\documentclass{iaus}

\usepackage{graphicx}

\usepackage{natbib}
\bibpunct{(}{)}{;}{a}{}{,} 

\usepackage{url}

\newcommand{\muas}[0]{\hbox{\rm $\mu$as}}

\newcommand{\ve}[1]{\mbox{\boldmath$#1$}}

\def\ov#1{{\overline{#1}}}
\def\arcsec{\hbox{$^{\prime\prime}$}}

\def\TDB{{\rm TDB}}
\def\TT{{\rm TT}}
\def\TCG{{\rm TCG}}
\def\TCB{{\rm TCB}}

\def\etal{{\it et al.}}
\def\aap{{\it A\&A}}

\begin{document}

\title[Relativistic Aspects of Rotational Motion]
{Relativistic aspects of rotational motion of celestial bodies}

\author[S.A. Klioner, E. Gerlach, M.H. Soffel]{S.A. Klioner, E. Gerlach, M.H. Soffel}

\affiliation{
Lohrmann Observatory, Dresden Technical University, 01062 Dresden, Germany
}

\pubyear{2009}
\volume{261}
\pagerange{1--9}
\setcounter{page}{1}
\jname{Relativity in Fundamental Astronomy:\\ Dynamics, Reference Frames, and Data Analysis}
\editors{S. Klioner, P. K. Seidelmann \& M. Soffel, eds.}

\maketitle

\begin{abstract}
  Relativistic modelling of rotational motion of extended bodies
  represents one of the most complicated problems of Applied
  Relativity. The relativistic reference systems of IAU (2000) give a
  suitable theoretical framework for such a modelling.  Recent
  developments in the post-Newtonian theory of Earth rotation in the
  limit of rigidly rotating multipoles are reported below.  All
  components of the theory are summarized and the results are
  demonstrated.  The experience with the relativistic Earth rotation
  theory can be directly applied to model the rotational motion of
  other celestial bodies. The high-precision theories of rotation of
  the Moon, Mars and Mercury can be expected to be of interest in the
  near future.
\end{abstract}

\maketitle

\firstsection
\section{Earth rotation and relativity}

Earth rotation is the only astronomical phenomenon which is observed
with very high accuracy, but is traditionally modelled in a Newtonian way.
Although a number of attempts to estimate and calculate the relativistic
effects in Earth rotation have been undertaken (e.g.,
\citet{Bizouardetal1992,BrumbergSimon2007} and reference therein) no
consistent theory has appeared until now. As a result the calculations
of different authors substantially differ from each other. Even the
way geodetic precession/nutation is usually taken into account is just
a first-order approximation and is not fully consistent with
relativity. On the other hand, the relativistic effects in Earth's
rotation are relatively large. For example, the geodetic precession
(1.9\arcsec\ per century) is about $3\times 10^{-4}$ of general
precession. The geodetic nutation (up to 200 \muas) is 200 times
larger than the goal accuracy of modern theories of Earth
rotation. One more reason to carefully investigate relativistic
effects in Earth rotation is the fact that the geodynamical
observations yield important tests of general relativity (e.g., the
best estimate of the PPN $\gamma$ using large range of angular
distances from the Sun comes from geodetic VLBI data) and it is
dangerous to risk that these tests are biased because of a 
relativistically flawed theory of Earth rotation.

Early attempts to model rotational motion of the Earth in a
relativistic framework \citep[see, for example,][]{Brumberg1972} made
use of only one relativistic references system to describe both
rotational and translational motions. That reference system was
usually chosen to be quite similar to the BCRS. This resulted in a
mathematically correct, but physically inadequate coordinate picture
of rotational motion. For example, from that coordinate picture a
prediction of seasonal LOD-variations with an amplitude of
about 75 microseconds has been put forward.

At the end of the 1980s a better reference system for modelling of
Earth rotation has been constructed, that after a number of
modifications and improvements has been adopted as GCRS in the IAU
2000 Resolutions. The GCRS implements the Einstein's equivalence
principle and represents a reference system in which the gravitational
influence of external matter (the Moon, the Sun, planets, etc.) is
reduced to tidal potentials. Thus, for physical phenomena occurring in
the vicinity of the Earth the GCRS represents a reference system, the
coordinates of which are, in a sense, as close as possible to
measurable quantities. This substantially simplifies the
interpretation of the coordinate description of physical phenomena
localized in the vicinity of the Earth. One important application of
the GCRS is modelling of Earth rotation.  The price to pay when using
GCRS is that one should deal not only with one relativistic reference
system, but with several reference systems, the most important of
which are the BCRS and the GCRS. This makes it necessary to clearly and
carefully distinguish between parameters and quantities defined in the
GCRS and those defined in the BCRS.

\section{Relativistic equations of Earth rotation}

The model which is used in this investigation was discussed and
published by \citet{Klioneretal2001}.
Let us, however, repeat these equations once again
not going into physical details.
The post-Newtonian equations of motion
(omitting numerically negligible terms as explained in 
\citet{Klioneretal2001}) read
\begin{eqnarray}\label{reduced-eqm}
{d\over dT}\,\left(C^{ab}\,\omega^b\right)&=&
\sum_{l=1}^\infty\ {1 \over l!}\
\varepsilon_{abc}\, M_{bL}\, G_{cL}
+L^a(\ve{C},\ve{\omega},\ve{\Omega}_{\rm iner}),
\end{eqnarray}

\noindent
where $T=\TCG$, $\ve{C}=C^{ab}$ is the post-Newtonian tensor of inertia and
$\ve{\omega}=\omega^a$ is the angular velocity of the post-Newtonian
Tisserand axes \citep{Klioner1996}, $M_{L}$ are the multipole
moments of the Earth's gravitational field defined in the GCRS,
$G_{L}$ are the multipole moments of the external tidal
gravito-electric field in the GCRS. In the simplest situation (a
number of mass monopoles) $G_L$ are explicitly given by Eqs. (19)--(23) of
\citet{Klioneretal2001}.

The additional torque $L^a$ depends on $\ve{C}$, $\ve{\omega}$, as
well as on the angular velocity $\ve{\Omega}_{\rm iner}$ describing
the relativistic precessions (geodetic, Lense-Thirring and Thomas
precessions). The definition of $\ve{\Omega}_{\rm iner}$ can be found,
e.g., in \citet{Klioneretal2001}. A detailed discussion of $L^a$, its
structure and consequences will be published elsewhere
\citep{Klioneretal2009}.

The model of rigidly rotating multipoles \citep{Klioneretal2001}
represents a set of formal mathematical assumptions that make the
general mathematical structure of Eqs. (\ref{reduced-eqm}) similar to
that of the Newtonian equations of rotation of a rigid body:

\begin{eqnarray}
\label{ovCij}
C^{ab}&=&P^{ac}\,P^{bd}\,\ov{C}^{cd},\quad \ov{C}^{cd}={\rm const}
\\
\label{ovM}
M_{a_1a_2\dots a_l}&=&P^{a_1b_1}\,P^{a_2b_2}\dots P^{a_lb_l}\,
\ov{M}_{b_1b_2\dots b_l},\quad
\ov{M}_{b_1b_2\dots b_l}={\rm const},
\quad l\ge2,
\end{eqnarray}
\noindent
where the orthogonal matrix $P^{ab}(T)$ is assumed to be related to
the angular velocity $\omega^a$ used in (\ref{reduced-eqm}) as

\begin{equation}\label{omegai-Pij}
\omega^a={1\over2}\,\varepsilon_{abc}\,P^{db}(T)\,{d\over dT}\,P^{dc}(T).
\end{equation}

\noindent
The meaning of these assumptions is that both the tensor of inertia
$C^{ab}$ and the multipole moments of the Earth's gravitational field
$M_L$ are ``rotating rigidly'' and that their rigid rotation is
described by the same angular velocity $\omega^a$ that appears in the
post-Newtonian equations of rotational motion. It means that in a
reference system obtained from the GCRS by a time-dependent 
rotation of spatial axes
both the tensor of inertia and the multipole moments of the Earth's
gravitational field are constant.

No acceptable definition of a physically rigid body exists in General
Relativity. The model of rigidly rotating multipoles represent a
minimal set of assumptions that allows one to develop the
post-Newtonian theory of rotation in the same manner as one usually
does within Newtonian theory for rigid bodies. In the model of rigidly
rotating multipoles only those properties of Newtonian rigid bodies
are saved which are indeed necessary for the theory of rotation.  For
example, no assumption on local physical properties (``local
rigidity'') is made. It has not been proved as a theorem, but it is
rather probable that no physical body can satisfy assumptions
(\ref{ovCij})--(\ref{omegai-Pij}).  The assumptions of the model of
rigidly rotating multipoles will be relaxed in a later stage of the
work when non-rigid effects are discussed. 

\section{Post-Newtonian equations of rotational motions 
in numerical computations}
 
Looking at the post-Newtonian equations of motion
(\ref{reduced-eqm})--(\ref{omegai-Pij}) one can formulate several
problems to be solved before the equations can be used in numerical
calculations:

\begin{list}{}{}
\item[A.]\ How to parametrize the matrix $P^{ab}$?
\item[B.]\ How to compute $M_L$ from the standard models of the Earth's
  gravity field?
\item[C.]\ How to compute $G_L$ from a solar system ephemeris?
\item[D.]\ How to compute the torque $\varepsilon_{abc}\, M_{bL}\, G_{cL}$
out of $M_L$ and $G_L$?
\item[E.]\ How to deal with different time scales (\TCG, \TCB, \TT, \TDB) 
appearing in the equations of motion, solar system ephemerides, used models
of Earth gravity, etc.?
\item[F.]\ How to treat the relativistic scaling of various
parameters when using \TDB\ and/or \TT\ instead of \TCB\ and \TCG? 
\item[G.]\ How to find relativistically meaningful numerical values for 
the initial conditions and various parameters?
\end{list}

These questions are discussed below.

\section{Relativistic definitions of the angles}

One of the tricky points is the definition of the angles describing
the Earth orientation in the relativistic framework. Exactly as in 
\citet{bretagnonetal1997,Bretagnonetal1998} we first define the
rotated BCRS coordinates $(x,y,z)$ by two constant rotations of the
BCRS as realized by the JPL's DE403:
\begin{equation}
\left(\matrix{x\cr y\cr z}\right)=R_x(23^\circ 26^\prime 21.40928^{\prime\prime})\,
R_z(-0.05294^{\prime\prime})\,\left(\matrix{x\cr y\cr z}\right)_{\rm DE403}.
\end{equation}
\noindent
Then the IAU 2000 transformations between BCRS and GCRS are applied to
the coordinates $(t,x,y,z)$, $t$ being \TCB, to get the corresponding
GCRS coordinates $(T,X,Y,Z)$. The spatial
coordinates $(X,Y,Z)$ are then rotated by the time-dependent 
matrix $P^{ij}$ to get the
spatial coordinates of the terrestrial reference system
$(\xi,\eta,\zeta)$. The matrix $P^{ij}$ is then 
represented as a product of three orthogonal matrices:
\begin{equation}
\left(\matrix{\xi\cr \eta\cr \zeta}\right)=
R_z(\phi)\,R_x(\omega)\,R_z(\psi)\,\left(\matrix{X\cr Y\cr Z}\right).
\end{equation}
\noindent
The angles $\phi$, $\psi$ and $\omega$ are used to parametrize the
orthogonal matrix $P^{ab}$ and therefore, to define the orientation of
the Earth orientation in the GCRS. The meaning of the terrestrial
system $(\xi,\eta,\zeta)$ here is the same as in
\citet{bretagnonetal1997}: this is the reference system in which we
define the harmonic expansion of the gravitational field with the
standard values of potential coefficients $C_{lm}$ and $S_{lm}$.

\section{STF model of the torque}

The relativistic torque requires computations with symmetric and trace-free cartesian (STF)
tensors $M_L$ and $G_L$. For this project special numerical algorithms for numerical
calculations have been developed. The detailed algorithms and their
derivation will be published elsewhere. Let us give here only the most
important formulas. For each $l$ the component $D_a=\varepsilon_{abc}\,
M_{bL-1}\, G_{cL-1}$ of the torque in the right-hand side of Eq.
(\ref{reduced-eqm}) can be computed as ($A_l=4\,l\,\pi\,l!/(2l+1)!!$,
$a^+_{lm}=\sqrt{l(l+1)-m(m+1)}\ $)
\begin{eqnarray}
\label{D1}
D_1&=&{1\over A_l}\,\Biggl(
\sum_{m=0}^{l-1}a^+_{lm}\,
\left(
-{\cal M}^R_{lm}\,{\cal G}^I_{l,m+1}
+{\cal M}^I_{l,m+1}\,{\cal G}^R_{lm}
\right)
\nonumber\\
&&
\qquad
+\sum_{m=1}^{l-1}a^+_{lm}\,
\left(
{\cal M}^I_{lm}\,{\cal G}^R_{l,m+1}
-{\cal M}^R_{l,m+1}\,{\cal G}^I_{lm}
\right)
\Biggr),
\\
\label{D2}
D_2&=&{1\over A_l}\,\Biggl(
\sum_{m=0}^{l-1}a^+_{lm}\,
\left(
-{\cal M}^R_{lm}\,{\cal G}^R_{l,m+1}
+{\cal M}^R_{l,m+1}\,{\cal G}^R_{lm}
\right)
\nonumber\\
&&
\qquad
+\sum_{m=1}^{l-1}a^+_{lm}\,
\left(
-{\cal M}^I_{lm}\,{\cal G}^I_{l,m+1}
+{\cal M}^I_{l,m+1}\,{\cal G}^I_{lm}
\right)
\Biggr),
\\
\label{D3}
D_3&=&{2\over A_l}\,\sum_{m=1}^l\,m\left({\cal M}^I_{lm}\,{\cal G}^R_{lm}-{\cal M}^R_{lm}\,{\cal G}^I_{lm}\right).
\end{eqnarray}
\noindent
The coefficients ${\cal G}^R_{lm}$ and ${\cal G}^I_{lm}$
characterizing the tidal field can be computed from Eqs. (19)--(23) of
\citet{Klioneretal2001} as explicit functions of the parameters of
the solar system bodies: their masses, positions, velocities and
accelerations. A Fortran code to compute ${\cal G}^R_{lm}$ and ${\cal
  G}^I_{lm}$ for $l<7$ and $0\le m\le l$ has been generated
automatically with a specially written software package for {\it
  Mathematica}. It is possible to develop a sort of recursive
algorithm to compute ${\cal G}^R_{lm}$ and ${\cal G}^I_{lm}$ for any
$l$ similar to the corresponding algorithms for, e.g., Legendre
polynomials.

The coefficients ${\cal M}^R_{lm}$ and ${\cal M}^I_{lm}$
characterizing the gravitational field of the Earth can be computed as
\begin{eqnarray}
\label{M_l0}
{\cal M}^R_{l0}&=&{l!\over (2l-1)!!}\,\left({4\pi\over 2l+1}\right)^{1/2}\,M_E\,R_E^l\,C_{l0},
\\
{\cal M}^R_{lm}&=&(-1)^m\,{1\over 2}\,{l!\over (2l-1)!!}\,\left({4\,\pi\over 2l+1}\,{(l+m)!\over (l-m)!}\right)^{1/2}\,M_E\,R_E^l\,C_{lm},\ 1\le m \le l,
\\
{\cal M}^I_{lm}&=&(-1)^{m+1}\,{1\over 2}\,{l!\over (2l-1)!!}\,\left({4\,\pi\over\,2l+1}\,{(l+m)!\over (l-m)!}\right)^{1/2}\,M_E\,R_E^l\,S_{lm}, \ 1\le m \le l,
\end{eqnarray}

\noindent
where $M_E$ is the mass of the Earth, $R_E$ its radius, $C_{lm}$ and
$S_{lm}$ are usual potential coefficients of the Earth's gravitational
field. If only Newtonian terms are considered in the torque this
formulation with STF tensors is fully equivalent to the classical
formulation with Legendre polynomials
\citep[e.g.,][]{bretagnonetal1997,Bretagnonetal1998}. If the
relativistic terms are taken in account, the only known way to
express the torque is that with STF tensors.

\section{Time transformations}

An important aspect of relativistic Earth rotation theory is the
treatment of different relativistic time scales. 
The transformation between \TDB\ and \TT\ {\it at the geocenter}
(all the transformations in this Section are meant to be ``evaluated
at the geocenter'') are computed along the lines of Section 3 of
\citet{Klioner2008b}. Namely,
\begin{eqnarray}
\TT=&\TDB&+\Delta\TDB(\TDB),
\\
\TDB=&\TT&-\Delta\TT(\TT),
\\
\TCG=&\TCB&+\Delta\TCB(\TCB),
\\
\TCB=&\TCG&-\Delta\TCG(\TCG),
\end{eqnarray}

\noindent
so that
\begin{eqnarray}
\label{deltaTDB}
{d\Delta\TDB\over d\TDB}&=&A_{\TDB}+B_{\TDB}\,{d\Delta\TCB\over d\TCB},\\
\label{A-TDB}
A_{\TDB}&=&{L_B-L_G\over 1-L_B},
\\
\label{B-TDB}
B_{\TDB}&=&{1-L_G\over 1-L_B}=A_{\TDB}+1,
\\
\label{deltaTT}
{d\Delta\TT\over d\TT}&=&A_{\TT}+B_{\TT}\,{d\Delta\TCG\over d\TCG},\\
\label{A-TT}
A_{\TT}&=&{L_B-L_G\over 1-L_G},
\\
\label{B-TT}
B_{\TT}&=&{1-L_B\over 1-L_G}=1-A_{\TT},
\\
\label{deltaTCB}
{d\Delta\TCB\over d\TCB}&=&F(\TCB)={1\over c^2}\,\alpha(\TCB)+{1\over c^4}\,\beta(\TCB),
\\
\label{deltaTCG}
{d\Delta\TCG\over d\TCG}&=&{F(\TCG-\Delta\TCG)\over 1+F(\TCG-\Delta\TCG)},
\end{eqnarray}

\noindent
where the functions $\alpha$ and $\beta$ are given by Eqs. (3.3)--(3-4) of
\citet{Klioner2008b} and Eq. (\ref{deltaTCG}) represents a computational
improvement of Eq. (3.8) of \citet{Klioner2008b}.  Clearly, the derivatives 
$d\Delta\TCB / dTCB$ and $d\Delta\TCG / dTCG$
must be expressed as functions of \TDB\ and \TT, respectively, when
used in (\ref{deltaTDB})--(\ref{deltaTT}). 

The differential equations for $\Delta\TDB$ and $\Delta\TT$ are first
integrated numerically for the whole range of the used solar system
ephemeris (any ephemeris with DE-like interface can be used with the
code).  The initial conditions for $\Delta\TDB$ and $\Delta\TT$ are 
chosen according to the IAU 2006 Resolution defining \TDB: for
$JD_{\TT} = 2443144.5003725$ one has $JD_{\TDB} =
2443144.5003725-6.55\times10^{−5}/86400$ and vice versa.  The results
of the integrations for the pairs $\Delta\TDB$ and $d\Delta\TCB / dTCB$, 
and $\Delta\TT$ and $d\Delta\TT / dTT$ are stored
with a selected step in the corresponding time variable (\TDB\ for
$\Delta\TDB$ and its derivative, and \TT\ for $\Delta\TT$ and its
derivative). A cubic spline on the equidistant grid is then
constructed for each of these 4 quantities.  The accuracy of the
spline representation is automatically estimated using additional data
points computed during the numerical integration. These additional
data points lie between the grid points used for the spline and are
only used to control the accuracy of the spline.  The splines
precomputed and validated in this way are stored in files and read in
by the main code upon request. These splines are directly used for
time transformation during the numerical integrations of Earth
rotation.  Although this spline representation requires significantly
more stored coefficients than, for example, a representation with
Chebyshev polynomials with the same accuracy, the spline
representation has been chosen because of its extremely high
computational efficiency. More sophisticated representations may be
implemented in future versions of the code.

\section{Relativistic scaling of parameters}

Obviously, there are two classes of quantities entering
Eqs. (\ref{reduced-eqm})--(\ref{omegai-Pij}) that are defined in the
BCRS and GCRS and, therefore, naturally parametrized by \TCB\ and \TCG,
respectively. It is important to realize that the post-Newtonian
equations of motion are only valid if non-scaled time scales \TCG\ and
\TCB\ are used. If \TT\ and/or \TDB\ are needed, the equations should
be changed correspondingly.

The relevant quantities defined in the GCRS and parametrized by \TCG\ are:
(1) the orthogonal matrix $P^{ab}$ and quantities related to that matrix: 
angular velocity $\omega^a$ and
corresponding Euler angles $\varphi$, $\psi$ and $\omega$; (2) the tensor of
inertia $C^{ab}$; 
(3) the multipole moment of Earth's gravitational field
$M_L$. In principle, (a) $G_L$ and (b) $\Omega^a_{\rm iner}$ are also
defined in the GCRS and parametrized by \TCG, but these quantities are
computed using positions $\ve{x}_A$, velocities $\ve{v}_A$ and
accelerations $\ve{a}_A$ of solar system bodies. The orbital motion of
solar system bodies are modelled in BCRS and parametrized by \TCB\ or
\TDB. The definition of $G_L$ is conceived in such a way that
positions, velocities and accelerations of solar system bodies in the BCRS
should be taken at the moment of \TCB\ corresponding to the required
moment of \TCG\ with spatial location taken at the geocenter.
Let us recall that the transformation between \TCB\ and \TCG\ is a
4-dimensional one and require the spatial location of an event to be
known.

In all theoretical works aimed to derive and/or analyze the rotational
equations of motion in the GCRS one uses \TCG\ as coordinate time
scale parametrizing the equations. Although the natural time variable
for the equations of Earth rotation is \TCG, in principle, using a
corresponding re-parametrization any time scale (including \TCG, \TT,
\TCB\ and \TDB) can be used as independent time variable. Thus, simple
rescaling of the first and second derivatives of the angles entering
the equations of rotational motion should be applied to use \TT\
instead of \TCG:
\begin{eqnarray}
\label{dotalpha-TT}
{d\theta\over d\TCG}&=&(1-L_G)\,{d\theta\over d\TT},
\\
\label{ddotalpha-TT}
{d^2\theta\over d\TCG^2}&=&(1-L_G)^2\,{d^2\theta\over d\TT^2},
\end{eqnarray}
\noindent
where $\theta$ is any of the angles $\varphi$, $\psi$ and $\omega$
used in the equations of motion to parametrize the orientation of the
Earth. If \TDB\ is used as independent variable the corresponding formulas
are more complicated:
\begin{eqnarray}
\label{dotalpha-TDB}
{d\theta\over d\TCG}&=&(1-L_G)\,{\left(\left.{d\TT\over d\TDB}\right|_{\ve{x}_E}\right)}^{-1}\,
{d\theta\over d\TDB},
\\
\label{ddotalpha-TDB}
{d^2\theta\over d\TCG^2}&=&(1-L_G)^2\,\left(\left.{d\TT\over d\TDB}\right|_{\ve{x}_E}\right)^{-2}\,{d^2\theta\over d\TDB^2}
\nonumber
\\
&&
-(1-L_G)^2\,\left(\left.{d\TT\over d\TDB}\right|_{\ve{x}_E}\right)^{-3}\,
\left.{d^2\TT\over d\TDB^2}\right|_{\ve{x}_E}\,
{d\theta\over d\TDB},
\end{eqnarray}
\noindent
where the derivatives of \TT\ w.r.t. \TDB\ should be evaluated at the
geocenter (i.e., for $\ve{x}=\ve{x}_E$). These relations must be
substituted into the equations of rotation motion to replace the
derivatives of the angles $\varphi$, $\psi$ and $\omega$ w.r.t. \TCG\
as appear e.g., in Eqs. (7)--(9) of \citet{Bretagnonetal1998}. It is
clear that the parametrization with \TDB\ makes the equations more
complicated.

The values of the parameters naturally entering the equations 
of rotational motion must be interpreted
as unscaled (\TCB-compatible or \TCG-compatible) values. If scaled
(\TT-compatible or \TDB-compatible) values are used, the scaling must be
explicitly taken into account. The relativistic scaling of parameters 
read \citep[see, e.g.,][]{Klioner2008a}:
\begin{eqnarray}
&&GM_A^{\TT}=(1-L_G)\,GM_A^{\TCG},\ 
GM_A^{\TCG}=GM_A^{\TCB},
\nonumber
\\
&&\qquad\qquad\qquad\qquad\qquad\qquad\quad
GM_A^{\TDB}=(1-L_B)\,GM_A^{\TCB},\\
&&X^{\TT}=(1-L_G)\,X^{\TCG}, \  
x^{\TDB}=(1-L_B)\,x^{\TCB},\\
&&V^{\TT}=V^{\TCG}, \  v^{\TDB}=v^{\TCB},\\
&&A^{\TT}=(1-L_G)^{-1}\,A^{\TCG}, \  a^{\TDB}=(1-L_B)^{-1}\,a^{\TCB},
\end{eqnarray}
\noindent
where $GM_A$ is the mass parameter of a body, $x$, $v$, and $a$
are parameters represents spatial coordinates (distances), 
velocities and accelerations in the BCRS, respectively, 
while $X$, $V$, and $A$ are similar quantities in the GCRS. 

Now, considering the source of the numerical values of the 
parameters used in the equations of Earth rotation we can
see the following.

\begin{itemize}

\item[a.] The position $\ve{x}_A$, velocities $\ve{v}_A$,
  accelerations $\ve{a}_A$ and mass parameters $GM_A$ of the massive
  solar system bodies are taken from standard JPL ephemerides and are
  \TDB-compatible. 

\item[b.] The radius of the Earth comes together with the potential
  coefficients $C_{lm}$ and $S_{lm}$ from a model of the Earth's gravity field
  (e.g., GEMT3 was used in SMART). These values come from SLR and
  dedicated techniques like GRACE. GCRS with \TT-compatible quantities
  is used to process these data. Therefore, the values of the radius
  of the Earth is \TT-compatible. Obviously, $C_{lm}$ and $S_{lm}$ have
  the same values when used with any time scale. The mass parameter
  $GM_E$ of the Earth coming with the Earth gravity models is also
  \TT-compatible. 

\item[c.] From the definitions of ${\cal M}^R_{lm}$ and ${\cal
    M}^I_{lm}$ given above and formulas for $G_L$ given by
  Eqs. (19)--(23) of \citet{Klioneretal2001}, it is easy to see
  that the \TCG-compatible torque $F^a= \sum_{l=1}^\infty\ {1 \over
    l!}\ \varepsilon_{abc}\, M_{bL}\, G_{cL}$ can be computed using
  \TDB-compatible values of mass parameters $GM_A^\TDB$, positions
  $\ve{x}_A^\TDB$, velocities $\ve{v}_A^\TDB$ and accelerations
  $\ve{a}_A^\TDB$ of all external bodies, \TDB-compatible value of the
  mass parameter of the Earth $GM_E^\TDB$ and the value of the Earth's
  radius formally rescaled from \TT\ to \TDB\ as
  $R_E^\TDB=(1-L_B)\,(1-L_G)^{-1}\,R_E^\TT$. Denoting the resulting
  torque by $F^a_\TDB$, it can be seen that the \TCG-compatible value is
  $F^a_\TCG=(1-L_B)^{-1}\,F^a_\TDB$.

\item[d.] The values of the Earth's moments of inertia ${\cal A}_i$,
  $i=1,2,3$ can be represented as $G{\cal A}_i=GM_ER_E^2k_i$, where
  $k_i$ is a factor characterizing the distribution of matter
  inside the Earth. Clearly, the factors $k_i$ do not depend on the
  scaling. Therefore, the moments of inertia can be scaled as
\begin{equation}
\label{Ai-scaling}
{\cal A}_i^{\TT}=(1-L_G)^3\,{\cal A}_i^{\TCG}. 
\end{equation}

\end{itemize}

The last question is how to interpret the values of the moments of
inertia ${\cal A}_i=(A,B,C)$ and the initial conditions for the angles
$\varphi$, $\psi$ and $\omega$ and their derivatives given in
\citet{Bretagnonetal1998}. Obviously, the initial angles at J2000
are independent of the scaling. For the other parameters in question
it is not possible to clearly claim if the given values are
\TDB-compatible or \TT-compatible. Arguments in favor of both
interpretations can be given. A rigorous solution here is only
possible when all calculations leading to these quantities are
repeated in the framework of General Relativity. In this paper we
prefer to interpret the SMART values of ${\cal A}_i$, $\dot\varphi$,
$\dot\psi$ and $\dot\omega$ as being \TT-compatible.  Therefore, if \TDB\
is used as independent variable, the values of the derivatives should
be changed accordingly. For any of these angles one has
\begin{equation}
\label{derivatives-TT-TDB}
{d\alpha\over d\TDB}=
{\left(\left.{d\TT\over d\TDB}\right|_{\ve{x}_E}\right)}\,{d\alpha\over d\TT}
\,.
\end{equation}
\noindent
Thus, we have all tools to treat correctly the relativistic scaling of
all relevant parameters of the Earth rotation theory as well as
relativistic time scales. 

\section{Geodetic precession and nutation}
\label{Section-geodetic}

In the framework of our model geodetic precession and nutation are
taken into account in a natural way by including the additional torque
that depends on $\Omega^a_{\rm iner}$ in the equations of rotational
motion:
\begin{equation}
\label{L-geodetic}
L^a=\varepsilon _{abc}\, C^{bd}\, \omega^d \,\Omega_{\rm iner}^c  - 
\frac{d}{dT}\left({C^{ab}\,\Omega_{\rm iner}^b}\right).
\end{equation}

\noindent
The first term of the additional torque reflects the fact that the
GCRS of the IAU is defined to be kinematically non-rotating
\citep[see][]{SoffelEtAl2003}. The second term has been usually
hidden by the corresponding re-definition of the post-Newtonian spin
\citep{DSXIII,KlionerSoffel2000}. It can be demonstrated that this
second term must be explicitly taken into account to maintain the
consistency between dynamically and kinematically non-rotating
solutions. Further details will be published elsewhere
\citep{Klioneretal2009}. Using the additional torque $L^a$ in
Eq. (\ref{reduced-eqm}) is a rigorous way to take geodetic
precession/nutation into account.

\begin{figure}
\begin{center}
\includegraphics[angle=0,width=9cm]{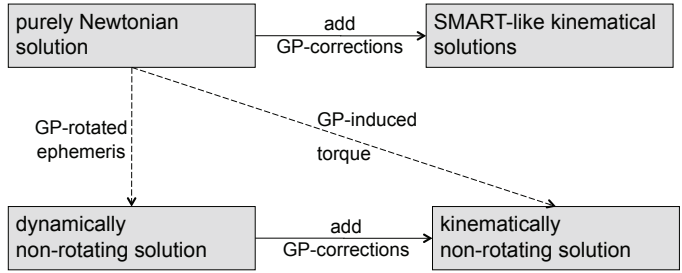}
\end{center}
\caption{Scheme of the two ways to obtain a kinematically non-rotating
  solution from a purely Newtonian one, and an illustration of the
  relation between the correct kinematically non-rotating solution and
  SMART-like kinematical solutions.  ``GP'' stands for geodetic
  precession/nutation. Each gray block represents a solution. A solid
  line means: add precomputed geodetic precession/nutation into a
  solution to get a new one. A dashed line means: recompute a solution
  with indicated change in the torque model.}
\label{DNR-KNR-schema}
\end{figure}

The standard way to account for geodetic precession/nutation that was used
up to now by a number of authors can be described as follows:
(1) solve the purely 
Newtonian equations of rotational motion and consider this solution as 
a relativistic one in a dynamically non-rotating version of the GCRS
and (2) add the precomputed geodetic precession/nutation to it. The
second step is fully correct since the geodetic precession/nutation is
by definition the rotation between the kinematically and dynamically
non-rotating versions of the GCRS and it can be precomputed since it
is fully independent of the Earth rotation. The inconsistency of the
first step comes from the fact that in the computation of the
Newtonian torque the coordinates of the solar system bodies are taken
from an ephemeris constructed in the BCRS. However, the dynamically
non-rotating version of the GCRS {\it rotates} relative to the BCRS
with angular velocity $\Omega^a_{\rm iner}(T)$. This means that the
BCRS coordinates of solar system bodies should be first rotated into
``dynamically non-rotating coordinates'' and only after that rotation
those coordinates can be used to compute the Newtonian torque.
For this reason this procedure does not lead to a correct solution
in the kinematically non-rotating GCRS (see Fig. \ref{DNR-KNR-schema}).
We will call such solutions in this paper ``SMART-like kinematical solutions''. 

On the other hand, there are two ways to obtain a correct
kinematically non-rotating solution: (1) use the torque given by 
Eq.~(\ref{L-geodetic}) in the equations of motion, (2) compute the
geodetic precession/nutation matrix, apply the geodetic
precession/nutation to the solar system ephemeris, integrate
(\ref{reduced-eqm}) without $L^a$ with the obtained rotated ephemeris
(the correct solution in a dynamically non-rotating version of the
GCRS is obtained in this step), apply the geodetic precession/nutation
matrix to the solution.  We have implemented both ways in our code and
checked explicitly that they give the same solution (to within about
0.001 \muas\ over 150 years). It is interesting to note that the
rotational matrix of geodetic precession/nutation (that is, the matrix
defining a rotation with the angular velocity $\Omega_{\rm iner}^a$)
cannot be parametrized by normal Euler angles. We have used therefore 
the quaternion representation for that matrix. 

\section{Overview of the numerical code}

A code in Fortran 95 has been written to integrate the post-Newtonian
equations of rotational motion numerically. The software is carefully
coded to avoid numerical instabilities and excessive round-off
errors. Two numerical integrators with dense output -- ODEX and
Adams-Bashforth-Moulton multistep integrator -- can be used for
numerical integrations.  These two integrators can be used to
crosscheck each other.  The integrations are automatically performed
in two directions -- forwards and backwards -- that allows one to directly
estimate the accuracy of the integration. The code is able to use any type
of arithmetic available with a given current hardware and
compiler. For a number of operations, which have been identified as
precision-critical, one has the possibility to use either the library
FMLIB \citet{Smith2001} for arbitrary-precision arithmetic or the
package DDFUN that uses two double-precision numbers to implement
quadrupole-precision arithmetic \citep{Bailey2005}. 
Our current baseline is to use ODEX with 80 bit
arithmetic. The estimated errors of numerical integrations after 150
years of integration are below $0.001$ \muas.

Several relativistic features have been incorporated into the code: (1)
the full post-Newtonian torque using the STF tensor machinery, (2)
rigorous treatment of geodetic precession/nutation as an additional
torque in the equations of motion, (3) rigorous treatment of time
scales (any of the four time scales -- \TT, \TDB, \TCB\ or \TCG\ --
evaluated at the geocenter can be used as the independent variable of
the equations of motion (\TCG\ being physically preferable for this
role), (4) correct relativistic scaling of constants and parameters.
All these ``sources of relativistic effects'' can be switched on and off
independently of each other.

In order to test our code and the STF-tensor formulation of the torque
we have coded also the classical Newtonian torque with Legendre
polynomials as described by \citet{bretagnonetal1997,Bretagnonetal1998} and
integrated our equations for 150 years with these two torque
algorithms. Maximal deviations between these two integrations were
0.0004 \muas\ for $\phi$, 0.0001 \muas\ for $\psi$, and 0.0002 \muas\
for $\omega$. This demonstrates both the equivalence of the two
formulations and the correctness of our code.

We have also repeated the Newtonian dynamical solution of SMART97
using the Newtonian torque, the JPL ephemeris DE403 and the same
initial values as in \citet{Bretagnonetal1998}. Jean-Louis
\citet{Simon2007} has provided us with the unpublished full version of
SMART97 (involving about 70000 Poisson terms for each of the three
angles). We have calculated the differences between that full SMART97
series and our numerical integration over 150 years. Analysis of the
results and a comparison to \citet{Bretagnonetal1998} have
demonstrated that our integrations reproduce SMART97 within the full
accuracy of the latter.

\section{Relativistic vs. Newtonian integrations}

\begin{figure}
\begin{center}
\includegraphics[angle=0,width=9cm]{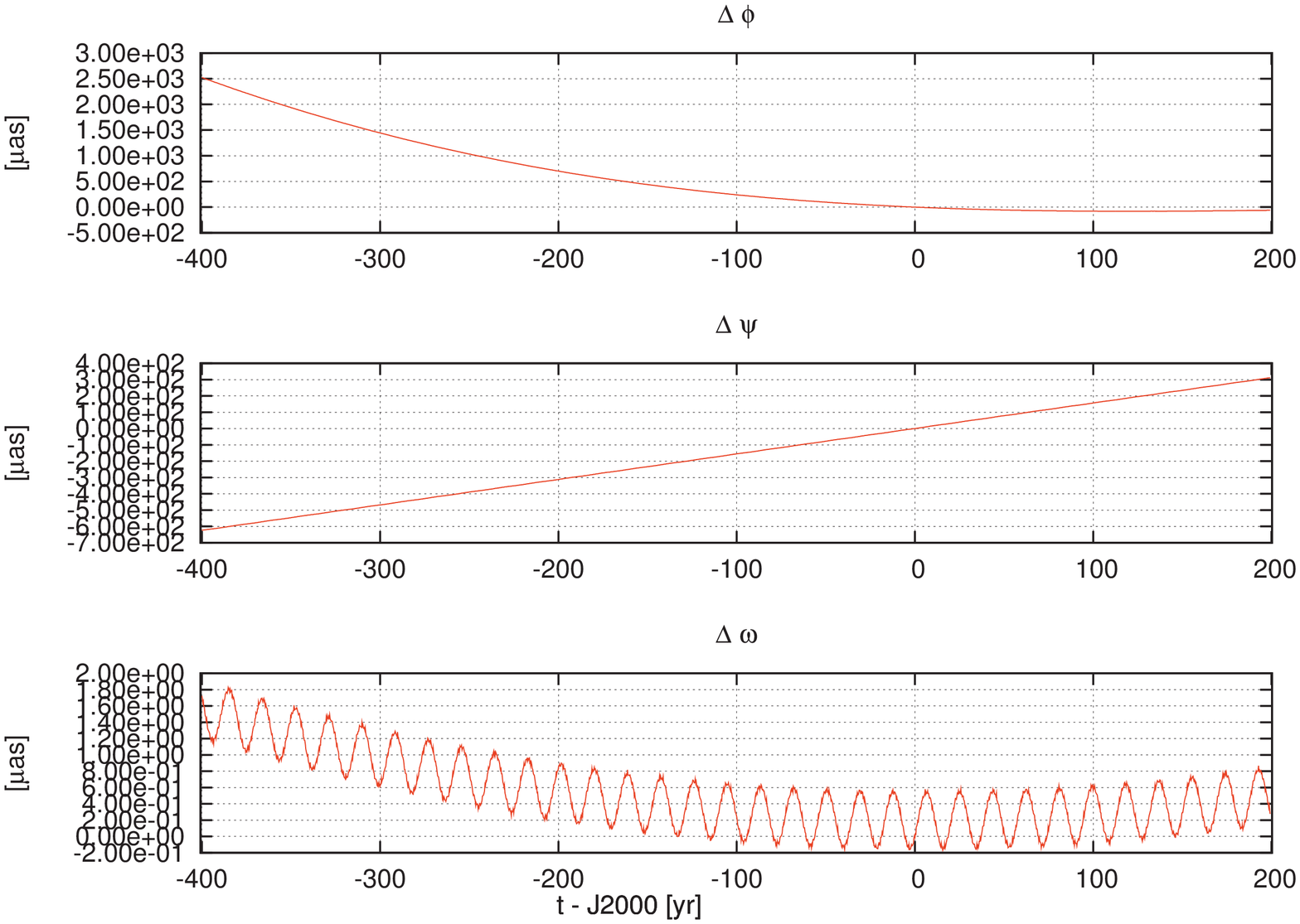}
\end{center}
\caption{The effect (in \muas) of the post-Newtonian torque.}
\label{Figure-pN-torque}
\end{figure}

\begin{figure}
\begin{center}
\includegraphics[angle=0,width=9cm]{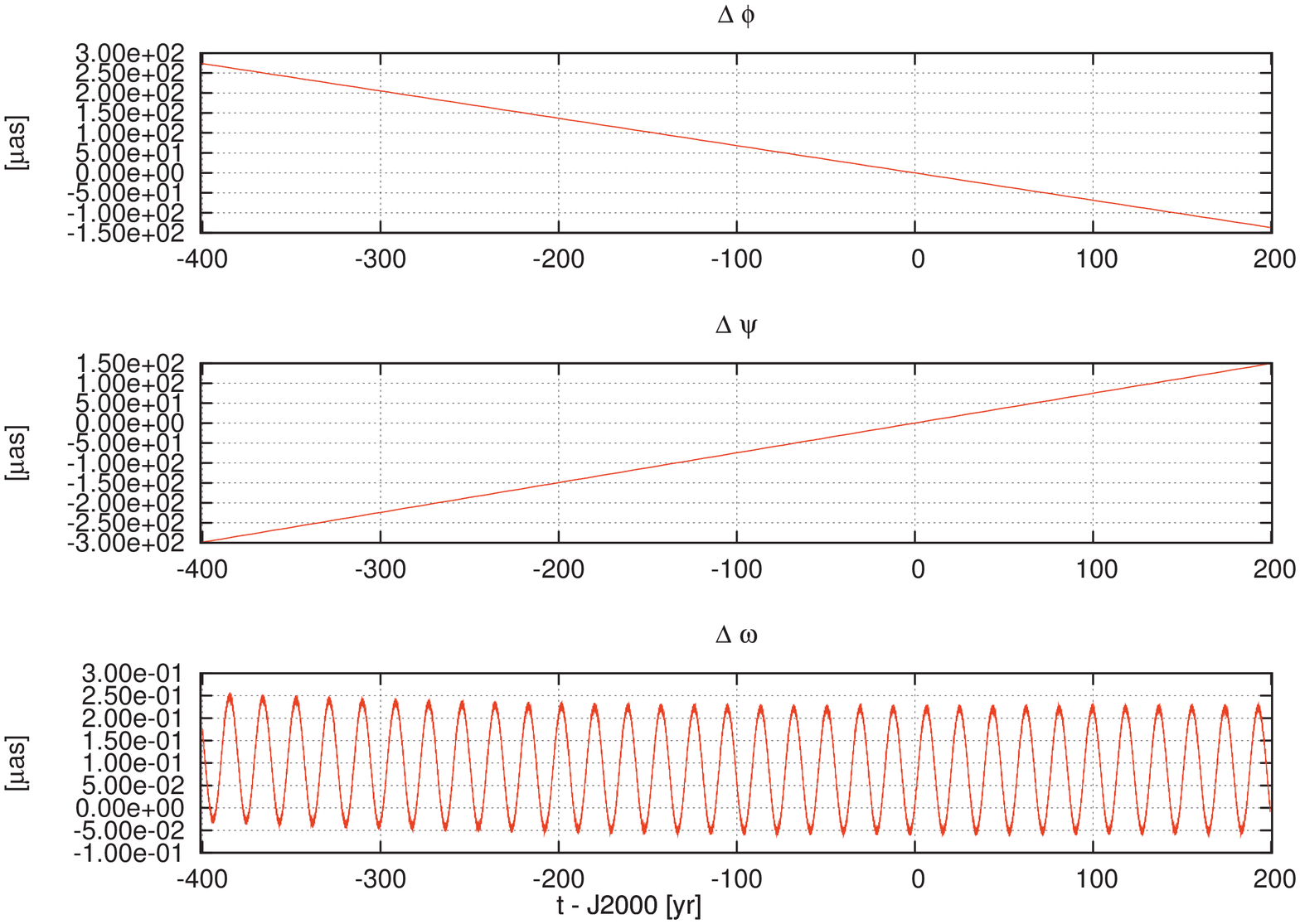}
\end{center}
\caption{The effect (in \muas) of the relativistic scaling and time scales.}
\label{Figure-scaling}
\end{figure}

\begin{figure}
\begin{center}
\includegraphics[angle=0,width=9cm]{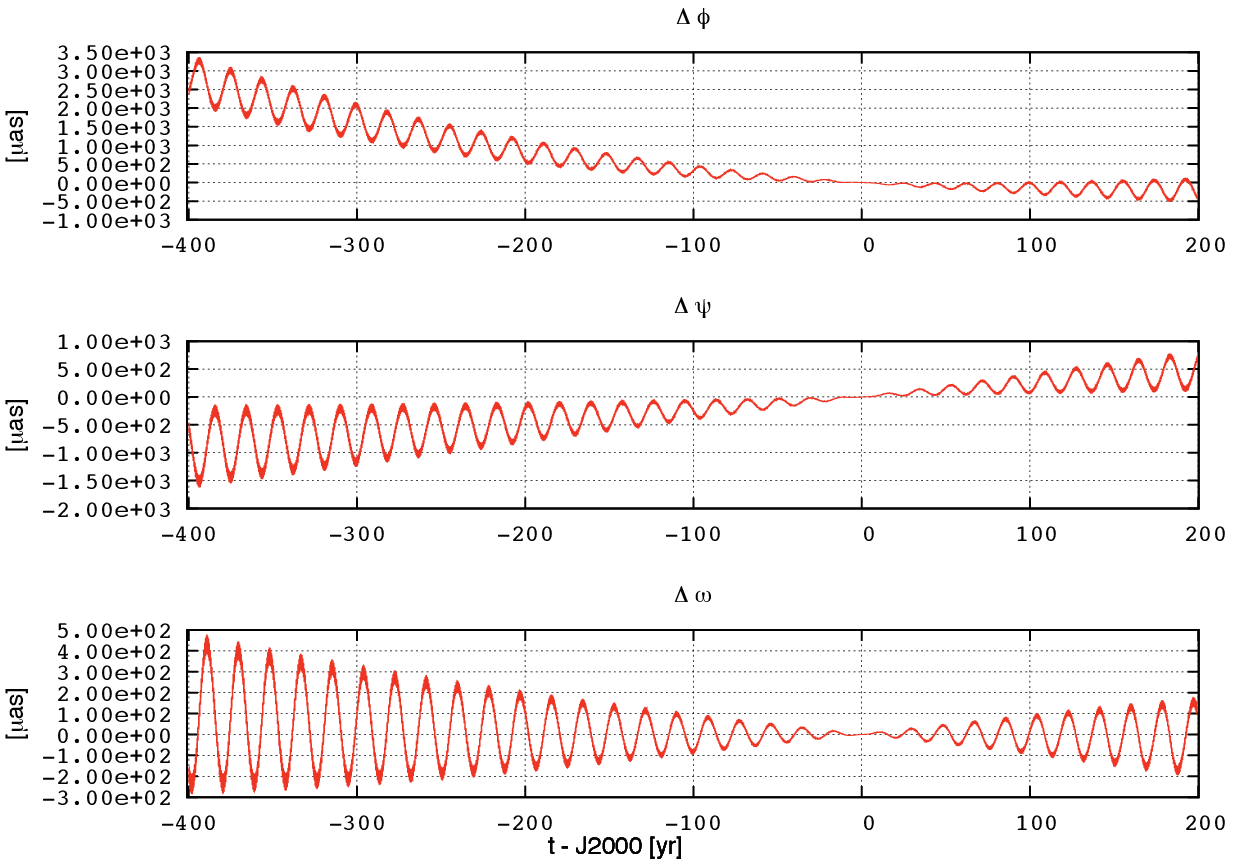}
\end{center}
\caption{Difference between a purely Newtonian integration rotated for
  geodetic precession/nutation in a SMART-like way (see Section
  \ref{Section-geodetic}) and our solution that includes all
  relativistic effects discussed here.}
\label{Figure-all}
\end{figure}

We have performed a series of numerical calculations comparing purely
Newtonian integration with integration where relativistic effects are
taken into account.  The same initial conditions and parameters that
we used to reconstruct the SMART97 solution were used for all
integrations (see below).  The results are illustrated on Figs.
\ref{KIN-diff}--\ref{Figure-all}. The difference between the
kinematical SMART97 solution and the consistent
kinematically-non-rotating solution obtained as described in Section
\ref{Section-geodetic} is shown in Fig. \ref{KIN-diff}.
Fig. \ref{Figure-pN-torque} shows the effects of the post-Newtonian
torque. The effects of the
relativistic scaling and time scales are depicted in
Fig. \ref{Figure-scaling}. Finally,
Fig. \ref{Figure-all} demonstrates the differences between a
SMART-like kinematical solution and our full post-Newtonian integration.
A detailed analysis of these results will be done elsewhere.

\begin{figure}
\begin{center}
\includegraphics[angle=0,width=9cm]{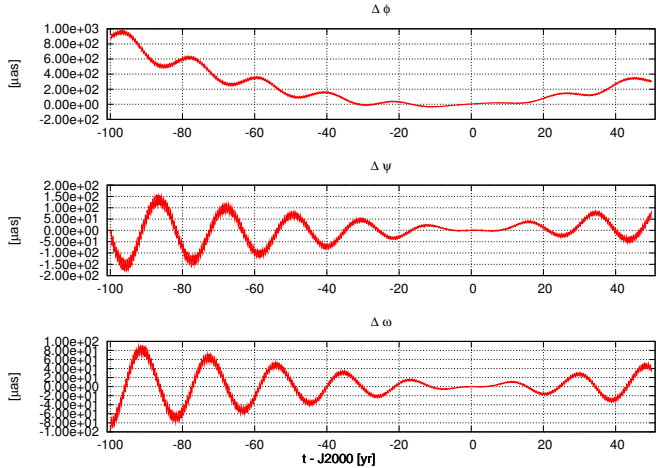}
\end{center}
\caption{Differences (in \muas) between the published
kinematical SMART97 solution and the correct
kinematically non-rotating solution (with post-Newtonian torques, 
relativistic scaling and time scaled neglected). }
\label{KIN-diff}
\end{figure}

To complete the consistent post-Newtonian theory of Earth rotation the
parameters (first of all, the moments of inertia of the Earth) should
be fitted to be consistent with the observed precession rate.  This
task will be discussed and treated in the near future.

\section{Relativistic effects in rotational bodies of other bodies}

The same numerical code can be applied to model the rotational motion
of other bodies. Especially, high-accuracy models of rotational motion
of the Moon, Mercury and Mars are of interest because of the planned
space missions to Mercury and Mars, and the expected improvements of
the accuracy of LLR. Most of the changes in the code are trivial and
concern the numerical values of the constants. One important
improvement of the code is necessary for the Moon: the figure-figure
interaction with the Earth must be taken into account. Using the STF
approach to compute the torque this task is not difficult.

The relativistic effects in the rotation of Moon, Mars and Mercury may
be significantly larger than in the rotation of Earth. In Table
\ref{Table-geodetic} the amplitudes of geodetic precession and
nutation are given for several solar system bodies. One can see 
the large effects for Mercury and Mars. Besides an early investigation
of \citet{BoisVokrouhlicky1995} suggests that the effects of the
relativistic torque for the Moon may attain 1 mas. Our approach
allows one to investigate the rotational motion of the Moon, Mars and
Mercury in a rigorous relativistic framework.

\begin{table}
\begin{center}
\begin{tabular}{lrr}
\hline
body & geodetic precession & geodetic nutation \\ 
     & [\,\arcsec\ per century\,] & [\,\muas\,] \\ 
\hline
Earth & 1.92  & 153 \\
Moon & 1.95 & 154 \\
Mercury & 21.43 & 5080 \\
Venus & 4.32 & 85 \\
Mars & 0.68 & 567
\end{tabular}
\end{center}
\caption{Magnitude of geodetic precession/nutation for various bodies}
\label{Table-geodetic}
\end{table}

\end{document}